\documentclass[aps,pra,reprint,showpacs,nofootinbib,superscriptaddress]{revtex4-1}
\input{Qcircuit}
\usepackage{graphicx, comment}
\usepackage{amsfonts}
\usepackage{amsmath,amssymb}
\usepackage{bm}
\usepackage{color}
\usepackage{epstopdf}
\usepackage{epsfig}
\usepackage{verbatim}
\usepackage{hyperref}
\hypersetup{
     colorlinks   = true,
     citecolor    = blue
}
\usepackage{lineno}
\usepackage[thicklines]{cancel}
\usepackage{color}
\usepackage{url}
    
{\rm }

\def\be{\begin{equation}}
 \def\ee{\end{equation}}
 \def\bea{\begin{eqnarray}}
 \def\eea{\end{eqnarray}}

\def\2{\frac{1}{2}}
\def\4{\frac{1}{4}}

\newcommand{\expect}[1]{\langle {#1} \rangle}

\begin{document}

\footnotetext{This manuscript has been authored by UT-Battelle, LLC, under Contract No. DE-AC0500OR22725 with the U.S. Department of Energy. The United States Government retains and the publisher, by accepting the article for publication, acknowledges that the United States Government retains a non-exclusive, paid-up, irrevocable, world-wide license to publish or reproduce the published form of this manuscript, or allow others to do so, for the United States Government purposes. The Department of Energy will provide public access to these results of federally sponsored research in accordance with the DOE Public Access Plan.}%

\title{Scalar Quantum Field Theories as a Benchmark for Near-Term Quantum Computers}

\author{K\"ubra Yeter-Aydeniz }
\affiliation{Department of Physics, Tennessee Technological University, Cookeville, TN 38505, USA}

\author{Eugene F.\ Dumitrescu}
\affiliation{Computational Sciences and Engineering Division, Oak Ridge National Laboratory,
  Oak Ridge, TN 37831, USA}

\author{Alex J.\ McCaskey}
\affiliation{Computer Science and Mathematics Division, Oak Ridge National Laboratory,
  Oak Ridge, TN 37831, USA}
 
\author{Ryan Bennink}
\affiliation{Computational Sciences and Engineering Division, Oak Ridge National Laboratory,
  Oak Ridge, TN 37831, USA}

\author{Raphael C.\ Pooser}
\email{pooserrc@ornl.gov}
\affiliation{Computational Sciences and Engineering Division, Oak Ridge National Laboratory,
  Oak Ridge, TN 37831, USA}
\affiliation{Department of Physics and Astronomy,  The University of Tennessee, Knoxville, TN 37996-1200, USA}

\author{George Siopsis}
\email{siopsis@tennessee.edu}
\affiliation{Department of Physics and Astronomy,  The University of Tennessee, Knoxville, TN 37996-1200, USA}

\date{\today}

\begin{abstract}
Quantum field theory (QFT) simulations are a potentially important application for noisy intermediate scale quantum (NISQ) computers. The ability of a quantum computer to emulate a QFT therefore constitutes a natural application-centric benchmark. Foundational quantum algorithms to simulate QFT processes rely on fault-tolerant computational resources, but to be useful on NISQ machines, error-resilient algorithms are required. Here we outline and implement a hybrid algorithm to calculate the lowest energy levels of the paradigmatic $1+1$--dimensional $\phi^4$ interacting scalar QFT. We calculate energy splittings and compare results with experimental values obtained on currently available quantum hardware. We show that the accuracy of mass-renormalization calculations represents a useful metric with which near-term hardware may be benchmarked. We also discuss the prospects of scaling the algorithm to full simulation of interacting QFTs on future hardware.
\end{abstract}

\maketitle

\section{Introduction}

The simulation of quantum field theories (QFTs) is expected to be a key application for quantum computers~\cite{feynman_simulating_1982,Lloyd1996} with digital~\cite{JordanLeePreskill2012} and analog~\cite{Marshall:2015mna} algorithms having already been proposed. The ability to efficiently simulate QFTs on quantum computers would allow us to make predictions with respect to the known Standard Model of physics.

Previous QFT algorithms were developed for universal fault-tolerant computers, operating at fixed logical error rates, in order to compute quantities to a known precision. On the other hand, in the Noisy Intermediate-Scale Quantum (NISQ)~\cite{Preskill2018quantumcomputingin} era, quantum simulation via noise-resilient hybrid quantum-classical algorithms~\cite{peruzzo_variational_2014}, hybrid digital-analog~\cite{arrazola_digital-analog_2016} algorithms, and fully analog computation~\cite{lloyd_quantum_1999,Martinez2016,clements_approximating_2017} have been developed as experimental computational methods in the presence of noisy environments. 
Recent results for QFTs, namely the Schwinger model~\cite{klco_quantum-classical_2018, Kokail2018}, showed that digital hybrid quantum-classical algorithms can be used to effectively simulate some aspects of QFTs on current quantum computers.

In this work, we reformulate previous QFT algorithms~\cite{JordanLeePreskill2012,Marshall:2015mna} as hybrid quantum-classical programs which are executed on available quantum hardware. We consider the paradigmatic relativistic quantum theory of scalar fields interacting via a quartic ($\phi^4$) potential. We discuss and demonstrate how to calculate the ground and excited state energies by hybrid variational algorithms, with IBM's Tokyo quantum processor~\cite{noauthor_quantum_2018} serving as example hardware. Our calculations illustrate how mass renormalization can be described by hybrid quantum-classical computations, and we also discuss phase transition to a symmetry-breaking phase. These results represent early progress towards quantum simulations of QFTs in the NISQ era. As such, our demonstrations, with limited depth (and width), can serve as practical high-level benchmarks for near-term technologies. 

The rest of the paper proceeds as follows. 
In Section~\ref{sec:set}, we review the $1+1$--dimensional $\phi^4$ scalar QFT on a lattice, mass renormalization, and the quantum discretization scheme. In Section~\ref{sec:sim}, we compare the quantum simulations performed on the IBM Tokyo superconducting quantum processor against classical numerical results. We conclude with a brief summary and discuss potential paths forward in  Section~\ref{sec:con}.

\section{Classical simulations}
\label{sec:set}

\subsection{The model}\label{sec:mod}
Let us begin by considering a real massive scalar field $\phi(x)$ in a single spatial dimension denoted by $x$. For numerical calculations, we discretize space into a finite set of lattice points along a chain of length $L$, and choose units so that the lattice spacing is $a=1$. Imposing periodic boundary conditions, the lattice coordinates $x$ are given by integers $x=0,1,\dots, L-1$ where $L$ is the full extent of the spatial dimension. In the chosen units, the continuum limit is obtained in the limit $L\to\infty$ as well as $m\to 0$.

Let $\pi$ be the conjugate momentum to $\phi$. We quantize the system by imposing the standard commutation relations
\be\label{eq1} [\phi (x) , \pi (x')] = i \delta_{xx'}~, \ee where we have also set $c, \hbar = 1$.
The model Hamiltonian, including the quartic interaction, is
\be \label{eq:H} H = \frac{1}{2} \sum_{x=0}^{L-1}  \left[ \pi^2(x) + (\nabla \phi(x))^2 + m_0^2 \phi^2(x) + \frac{\lambda}{12} \phi^4 (x) \right]~, \ee
where $\nabla\phi (x) = \phi(x+1) - \phi(x)$ is the finite difference gradient operator, $m_0$ is the bare mass of the field, and $\lambda$ is the interaction strength (of units mass-squared). The interaction term introduces quantum effects that result in a renormalized physical mass $m$ which differs from $m_0$. 

To account for the difference of the two mass (energy) parameters, we define the mass counter term by
\be\label{eq:3} \delta_m = m_0^2 - m^2~. \ee
The value of the counter term can be calculated analytically as a perturbative expansion in the coupling constant $\lambda$,  as will be discussed below. In the strong-coupling regime, $\delta_m$ can only be deduced in retrospect from the poles of correlators (Green functions) or the energy levels of the Hamiltonian. Here, we will discuss the former approach for perturbative analytic calculations, and the latter for numerical calculations and experimental results.

Using Eq.~\eqref{eq:3}, we can re-write the total Hamiltonian \eqref{eq:H} as $H=H_0+H_I$, with
\begin{eqnarray}
H_0 &=& \frac{1}{2} \sum_{x=0}^{L-1}  \left[ \pi^2(x) + (\nabla \phi(x))^2 + m^2 \phi^2(x)  \right]~, \nonumber\\
H_I &=& \sum_{x=0}^{L-1}  \left[ \frac{\delta_m}{2} \phi^2(x) + \frac{\lambda}{4!} \phi^4 (x) \right]~.
\label{Hamiltonian}\end{eqnarray}
We may expand the field $\phi$ and its conjugate momentum $\pi$ in terms of the canonical creation and annihilation operators as
\begin{eqnarray}
\label{eq:fields}
\phi(x) &=& \frac{1}{\sqrt{L}}\sum_k\frac{1}{\sqrt{2\omega(k)}}\left( a^\dagger(k) e^{-ik x} + a(k) e^{ik x} \right), \nonumber \\
\pi(x) &=& \frac{i}{\sqrt{L}}\sum_k\sqrt{\frac{\omega({k})}{2}} \left( a^\dagger(k) e^{-ik x} -a(k) e^{ik x} \right) 
\end{eqnarray}
where the momentum $k$ resides on the dual lattice ($k\in 2\pi\mathbb{Z}/L)$, $a^\dagger(k), a(k)$ are the plane wave creation and annihilation operators, respectively, and \be\label{eq4a} \omega^2 (k) = m^2 + 4  \sin^2\frac{k}{2} \ee is the free dispersion relation.  

The canonical commutation relations \eqref{eq1} lead to the standard commutation relations between the creation and annihilation operators of the different Fourier modes, \be [a({k}), a^\dagger({k'})] =\delta_{k k'}~,\ee with all other commutators vanishing. Transforming Eq.~\eqref{Hamiltonian} via the field expansion of Eq.~\eqref{eq:fields}, the free Hamiltonian is diagonalized,
\begin{equation}\label{eq3}
H_0 = \sum_{k} \omega({k}) a^\dagger(k) a(k)~,
\end{equation}
with its spectrum given by the free dispersion relation \eqref{eq4a}. We discard the zero-point energy, for simplicity. 


\subsection{Perturbative calculations}
Treating $H_I$ as a perturbation, we note that at zeroth order (i.e., ignoring $H_I$), $m$ is the mass gap, i.e., the difference in the energy levels between the ground state $|0\rangle$ and the first excited state $a^\dagger (0) |0\rangle$. Since $m$ is the physical mass, it should coincide with the gap at all orders in perturbation theory. Therefore, the mass gap should not receive any corrections at higher perturbative orders. This leads to a determination of the counter term $\delta_m$ (or, equivalently, the bare mass $m_0$) as a series expansion in the coupling constant $\lambda$.

\begin{figure}[ht!]
	\includegraphics[scale=0.65]{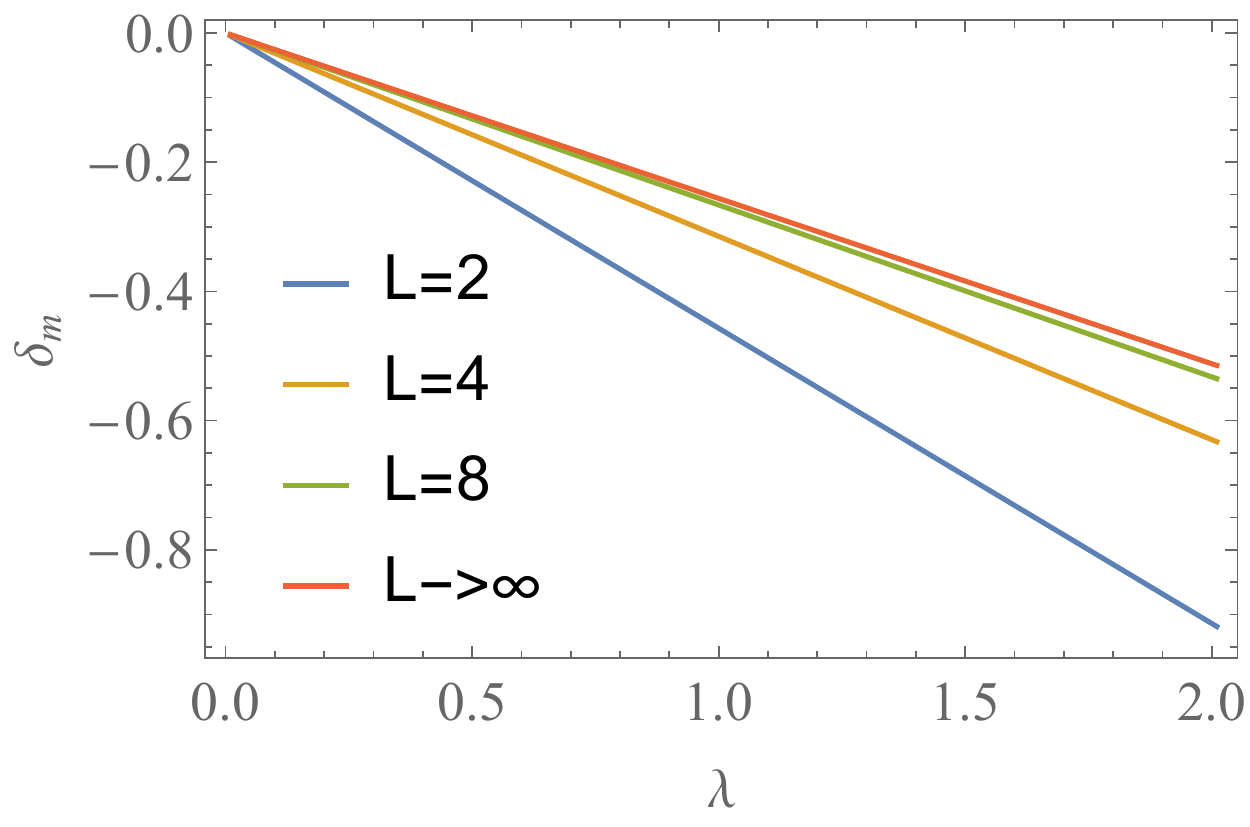}
	\includegraphics[scale=0.65]{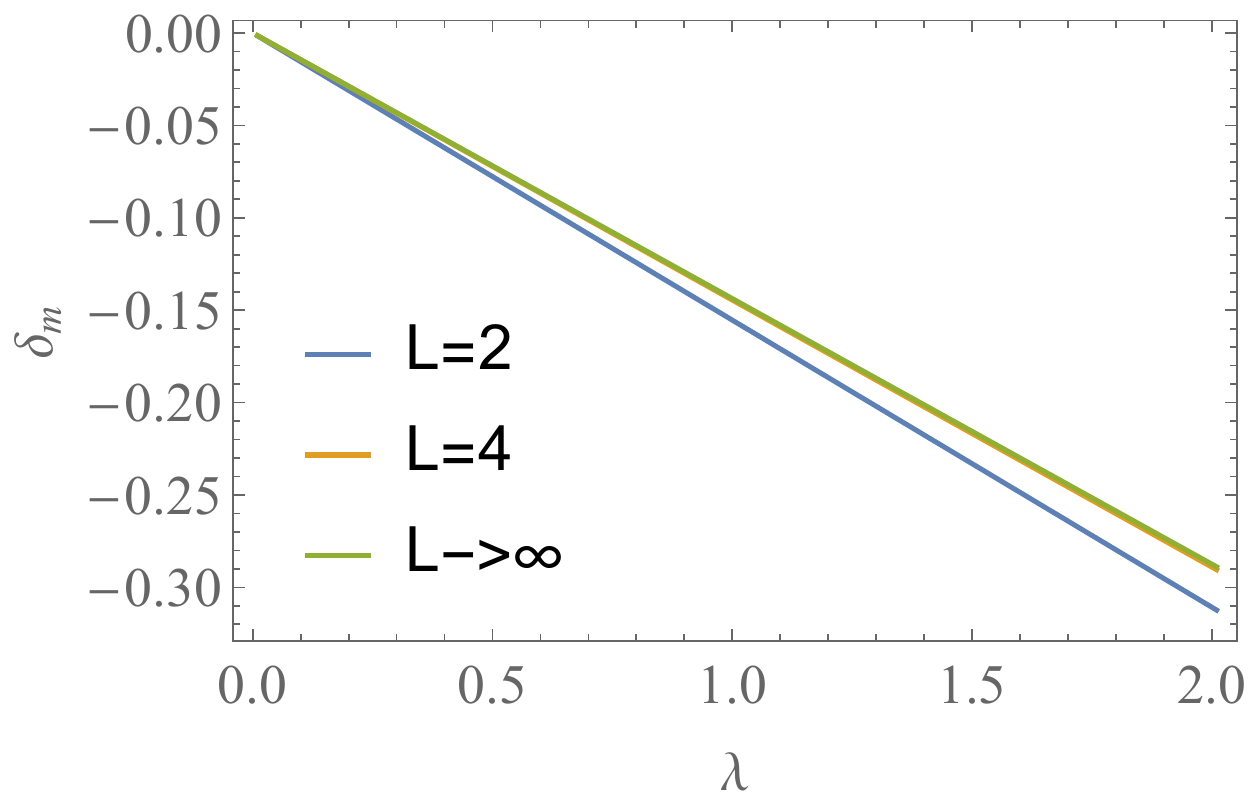}
\caption{The counter term, $\delta_m$, \textit{vs.}\ the interaction coefficient, $\lambda$, at first perturbative order for various values of $L$, including the continuum limit $L\to\infty$. The mass is chosen as $m^2 = 0.1$ (upper panel) and $m^2 = 1.5$ (lower panel). \label{fig:GS1}}
\end{figure}

At first perturbative order, after some algebra, we obtain a correction to the gap which is proportional to
\be \delta_m + \frac{\lambda}{4L} \sum_{k} \frac{1}{\omega(k)}~. \ee
Demanding that the correction vanish (in order for the gap to remain at the chosen value $m$), we deduce
\be \delta_m = - \frac{\lambda}{4L} \sum_{k} \frac{1}{\omega(k)}~. \ee
Thus, for a given physical mass $m$, we ought to choose the counter term
\be \delta_m = - \frac{\lambda}{4L} \sum_{k} \frac{1}{\omega(k)} + \mathcal{O} (\lambda^2) \label{eq:delta_m} \ee
In the continuum limit ($L\to\infty$, and also $m\to 0$ in units in which the lattice spacing is $a=1$), this becomes
\be \delta_m = - \frac{\lambda}{4} \int_{-\pi}^\pi \frac{dk}{\omega(k)} + \mathcal{O} (\lambda^2) = - \frac{\lambda}{8\pi} \log \frac{64}{m^2} + \dots \ee
Higher perturbative orders can be computed similarly.

Figure \ref{fig:GS1} shows the counter term $\delta_m$ as a function of the coupling constant $\lambda$ at first perturbative order for finite $L$ and the continuum limit ($L\to\infty$). We have chosen $m^2=0.1, 1.5$. Perturbation theory is valid at weak coupling, i.e., $\lambda /m^2 \lesssim 1$.

For fixed $\lambda$ in the continuum limit, the system approaches a critical point as $m^2\to 0$. Below that point ($m^2 < 0$), the system undergoes symmetry breaking. To approach criticality, we need to go beyond perturbation theory, because as $m^2 \to 0$, we have $\lambda/ m^2 \to \infty$.


\begin{figure}[ht!]
	\begin{center}
	\includegraphics[scale=0.65]{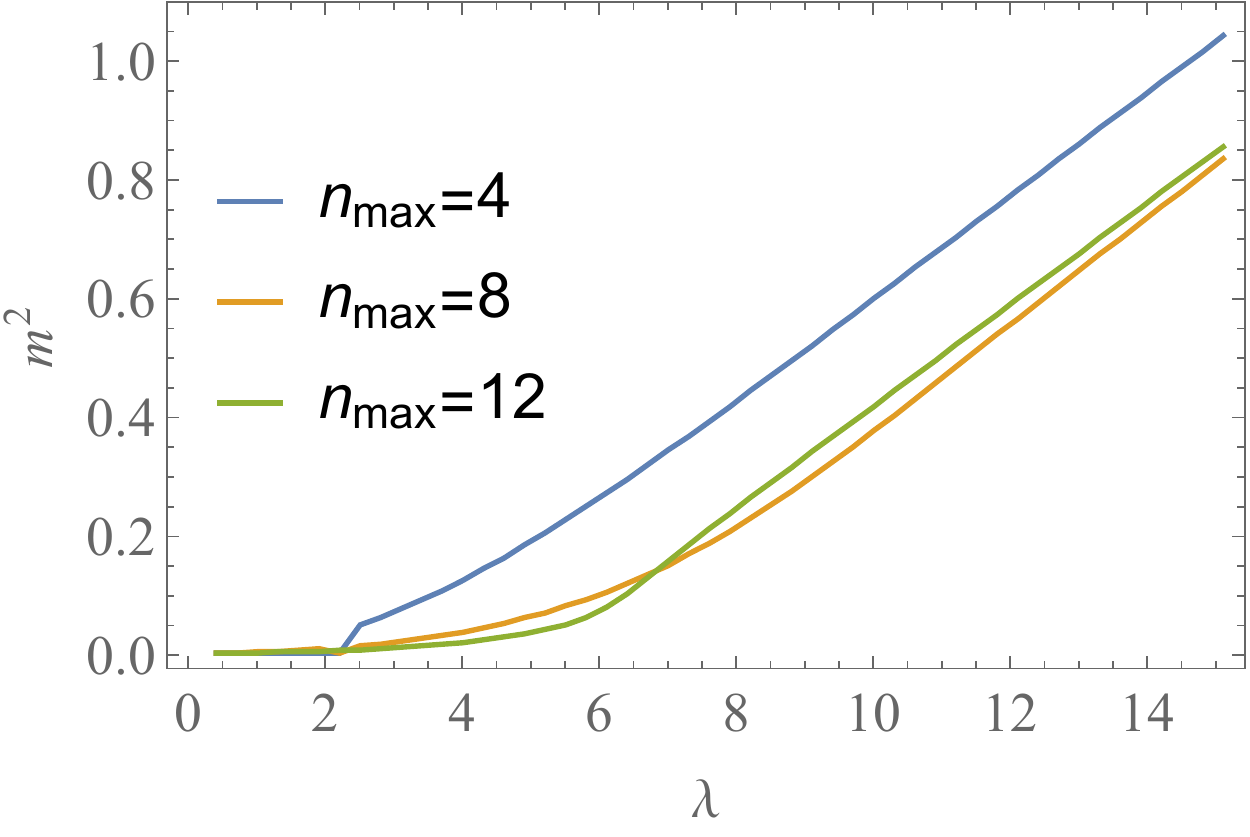}
	\end{center}
\caption{The mass gap \textit{vs.}\ the interaction coefficient $\lambda$ for $m_0^2=-1.5$, $L=2$, and Hilbert space cutoff $n_{\text{max}} =4, 8, 12$. 
}\label{fig:GS2}
\end{figure}

\begin{figure}[ht!]
	\includegraphics[scale=0.65]{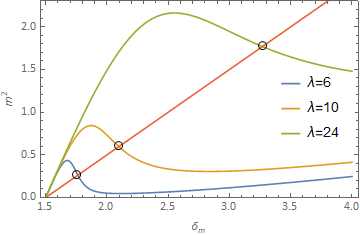}
	\includegraphics[scale=0.65]{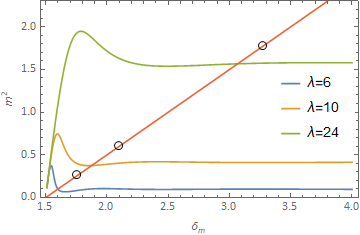}
	\includegraphics[scale=0.65]{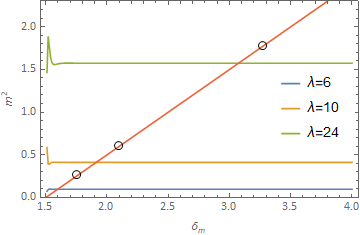}
\caption{The mass gap dependence on the parameter $\delta_m$ for $L=2$, $m_0^2=-1.5$, and for $\lambda =6, 10, 24$. The Hilbert space cutoff is $n_{\text{max}} =4, 8, 12$, top to bottom panels. Circles indicate the position of the counter term $\delta_m$ corresponding to the mass gap $m$ for $n_{\text{max}} =4$.}\label{gap_vs_deltam_2pts}
\end{figure}

\subsection{Numerical calculations}
\label{subsec:Class_Sim}
The physical mass, $m$, can be calculated by finding the energy difference between the first excited state and the ground state (mass gap $m=E_1-E_0$). However, $m$ is not a parameter that appears in the Hamiltonian \eqref{eq:H} of the system. The latter is parametrized by the bare mass $m_0$ and the coupling constant $\lambda$. If $m_0^2 >0$, then the eigenstates of the free Hamiltonian (setting $\lambda =0$) are well-defined, and one may use them to build the Hilbert space. Unfortunately, the physically relevant domain is near the critical line in the $(m_0^2,\lambda)$ plane, where $m_0^2 <0$ and the free Hamiltonian is ill-defined. One may, instead, use the free Hamiltonian $H_0$ (Eq.\ \eqref{Hamiltonian}) with mass parameter $m^2 >0$. The mass parameter need not be the physical mass; any parameter with $m^2 >0$ will do.
This is because the mass gap and other physical quantities should not depend on this parameter, and consequently the counter-term parameter $\delta_m$ can be chosen arbitrarily. However, due to the truncation of the Hilbert space of each mode (harmonic oscillator), which is necessary for numerical calculations, there is a dependence on $\delta_m$, albeit mild.

Let us truncate the Hilbert space of each mode to dimension $n_{\text{max}}$, so that the creation and annihilation operators become $n_{\text{max}}\times n_{\text{max}}$ matrices.
Figure \ref{fig:GS2} shows the error due to this truncation. Notice that $n_{\text{max}} = 8$ is already a good approximation to the continuum limit, being almost indistinguishable from the higher cutoff $n_{\text{max}} = 12$. 

Figure \ref{gap_vs_deltam_2pts} shows the dependence of the mass gap on the parameter $\delta_m$ for a fixed value of the bare mass parameter $m_0^2$ and various values of the coupling constant $\lambda$ and cutoff $n_{\text{max}}$ of the Hilbert space. As expected the dependence on $\delta_m$ is stronger for smaller cutoff $n_{\text{max}}$. For our calculations, we will choose the value of $\delta_m$ that satisfies Eq.\ \eqref{eq:3} with $m$ being the physical mass (counter term). The value of the counter term is indicated in Fig.\ \ref{gap_vs_deltam_2pts} by circles for cutoff value $n_{\text{max}} =4$, and compared with the corresponding values at higher cutoffs. Notice that, even though there is a strong dependence on $\delta_m$ for $n_{\text{max}} =4$, the values of the counter term (circles) are a reasonably good approximation to the true value of the counter term (obtained in the limit $n_{\text{max}} \to\infty$).

\begin{figure}[ht!]
	\begin{center}
	\includegraphics[scale=0.65]{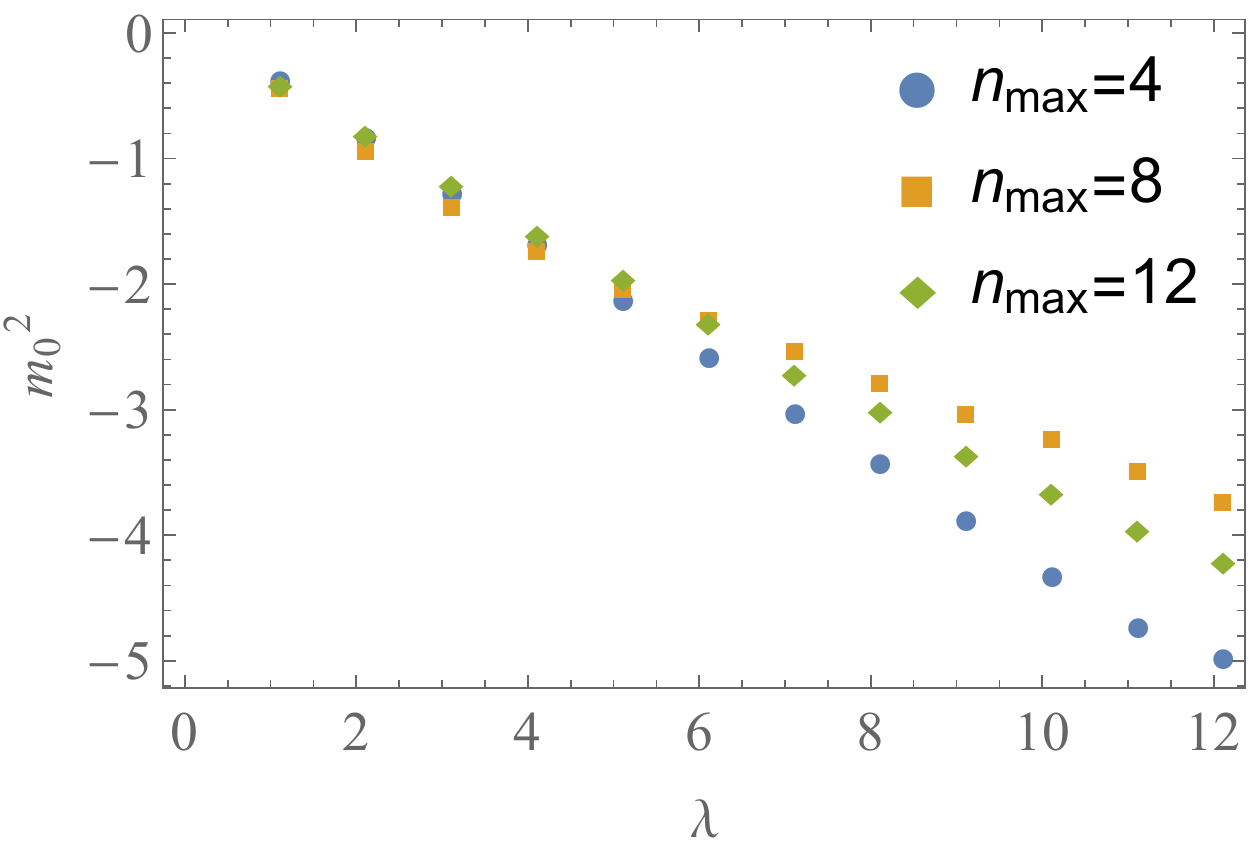}
	\includegraphics[scale=0.65]{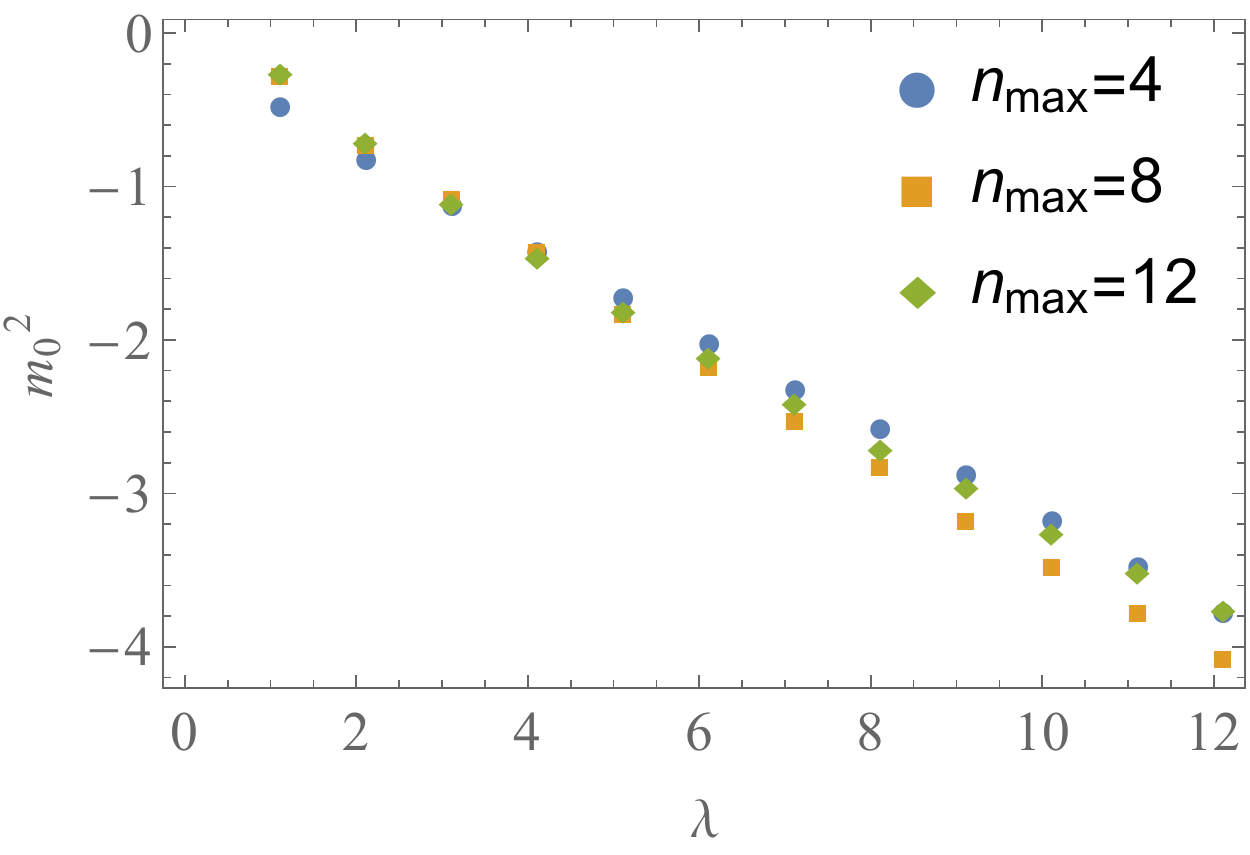}
	\includegraphics[scale=0.65]{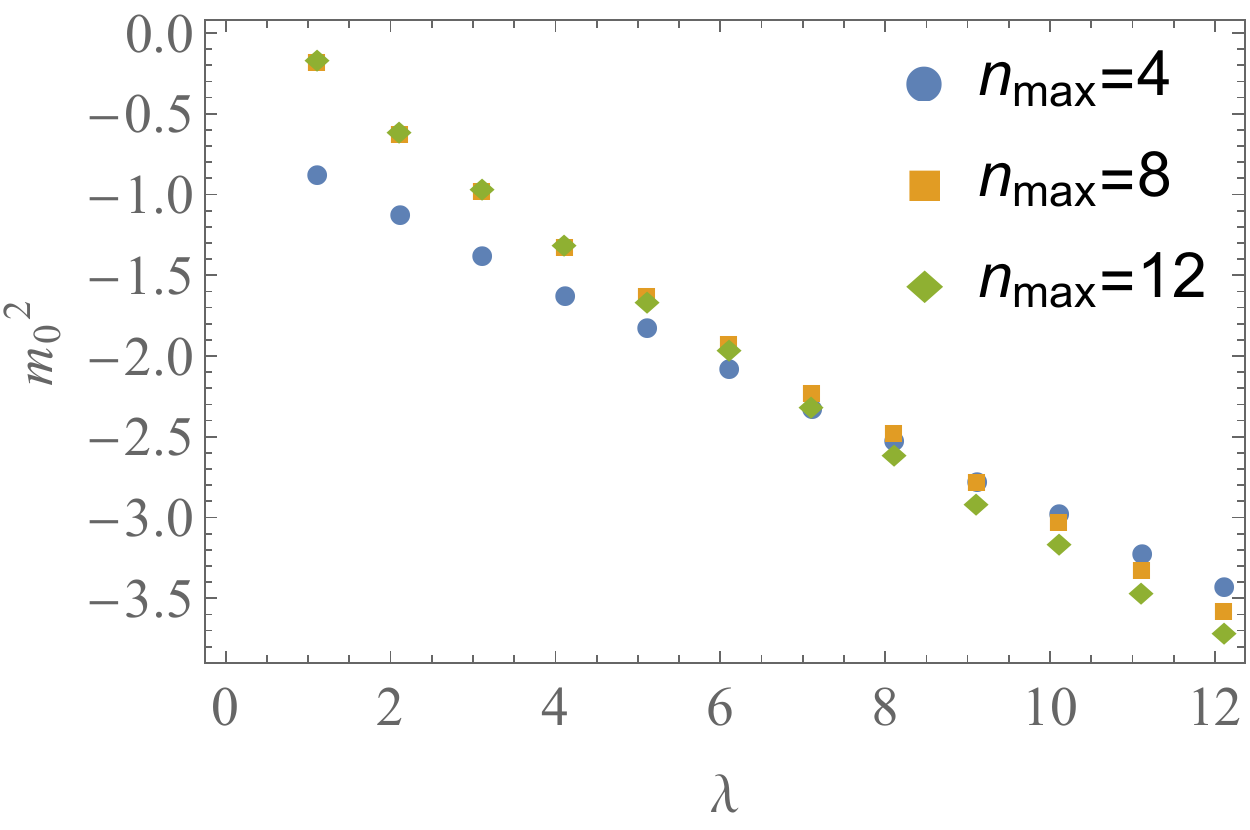}
	\end{center}
\caption{$m_0^2$ \textit{vs.}\ interaction coefficient $\lambda$ for $L=2$, mass gap $m^2= 0.1, 0.25, 0.5$, and cutoff $n_{\text{max}} =4 , 8, 12$.} \label{criticalcurve_2pts}
\end{figure}

\begin{figure}[ht!]
	\includegraphics[scale=0.65]{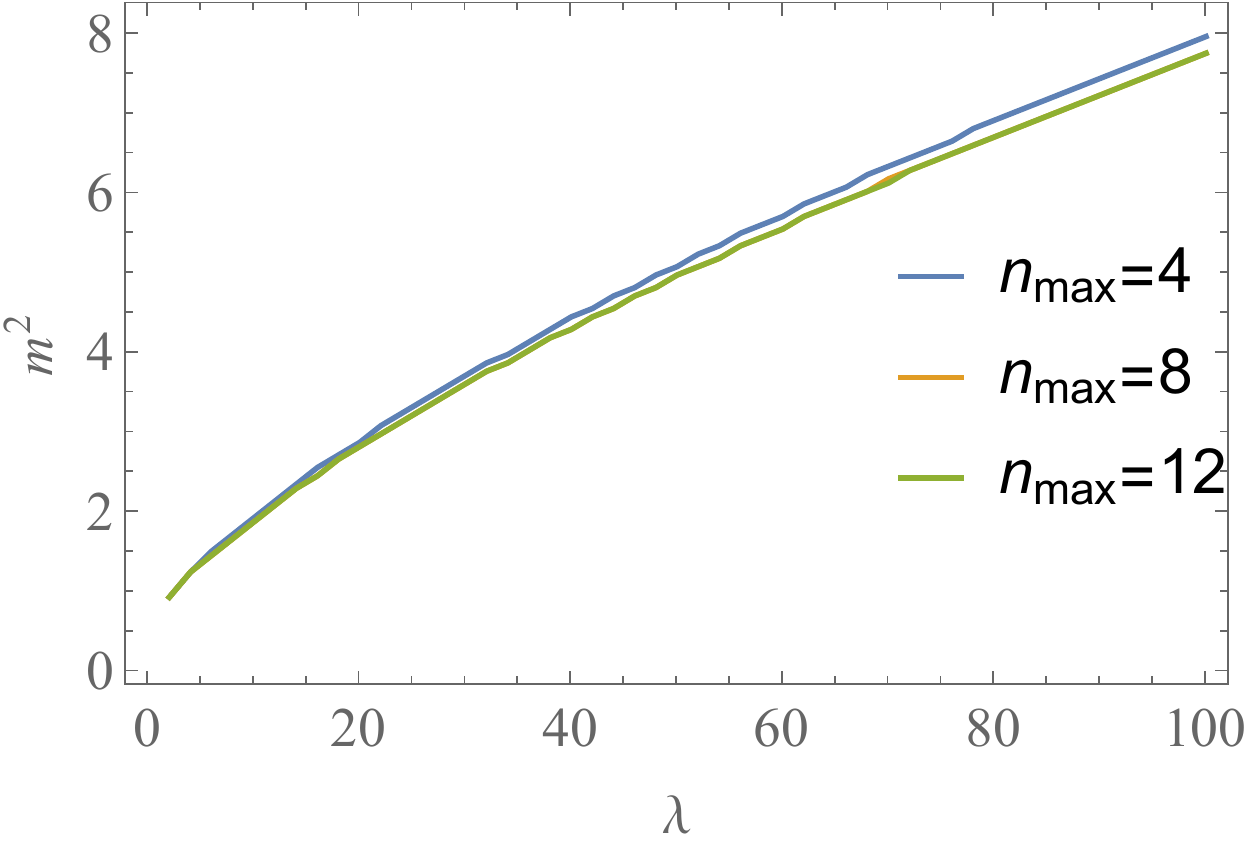}
\caption{The dependence of the mass gap on the interaction coefficient $\lambda$ for $m_0^2=0.5$, $L=2$, showing a weak dependence on the cutoff such that even the $n_{\text{max}} =4$ digitization is a reasonable approximation.
\label{fig:1}}
\end{figure}

As indicated above, the error due to the introduction of the cutoff $n_{max}$ increases as the dimensionless coupling constant $\lambda / m^2$ increases. This occurs as we approach the critical curve ($m^2 \to 0$). One can estimate the error due to the cutoff by comparing with a larger cutoff. Such a comparison is performed in Figure \ref{criticalcurve_2pts}. As expected, the error increases as $m^2 \to 0$. For comparison, Figure \ref{fig:1} shows that the error is negligible for positive values of $m_0^2$, so that even the digitization with cutoff $n_{\text{max}} =4$ is a reasonably good approximation for a wide range of the coupling constant $\lambda$. As seen in figure \ref{fig:1}, $n_{\text{max}} = 8, 12$ are indistinguishable.
[ht!]
As mentioned earlier, a phase transition is observed as the physical mass vanishes ($m^2 \to 0^+$) for particular $m_0^2$ and $\lambda$ values in the continuum limit. As we go past the critical line, we enter a phase in which we have symmetry breaking. There is, of course, no phase transition on a finite lattice, but the critical line in the $(m_0^2,\lambda)$ plane may still be approximated by taking the limit $m^2 \to 0$ on the finite lattice.  
Fig.\,\ref{criticalcurve_2pts} shows approximations of the critical curve for mass gap values close to zero ($m^2 = 0.1, 0.25, 0.5$). Remarkably, even with a Hilbert space cutoff $n_{\text{max}} =4$, we obtain reasonably good approximations. As expected, the approximation is not as good for the smallest value of the mass gap ($m^2 = 0.1$). As we increase the coupling constant $\lambda$, the approximation diverges from its true value, because we need a larger cutoff $n_{\text{max}}$. 
\begin{figure}[ht!]
	\begin{center}	
	\includegraphics[scale=0.65]{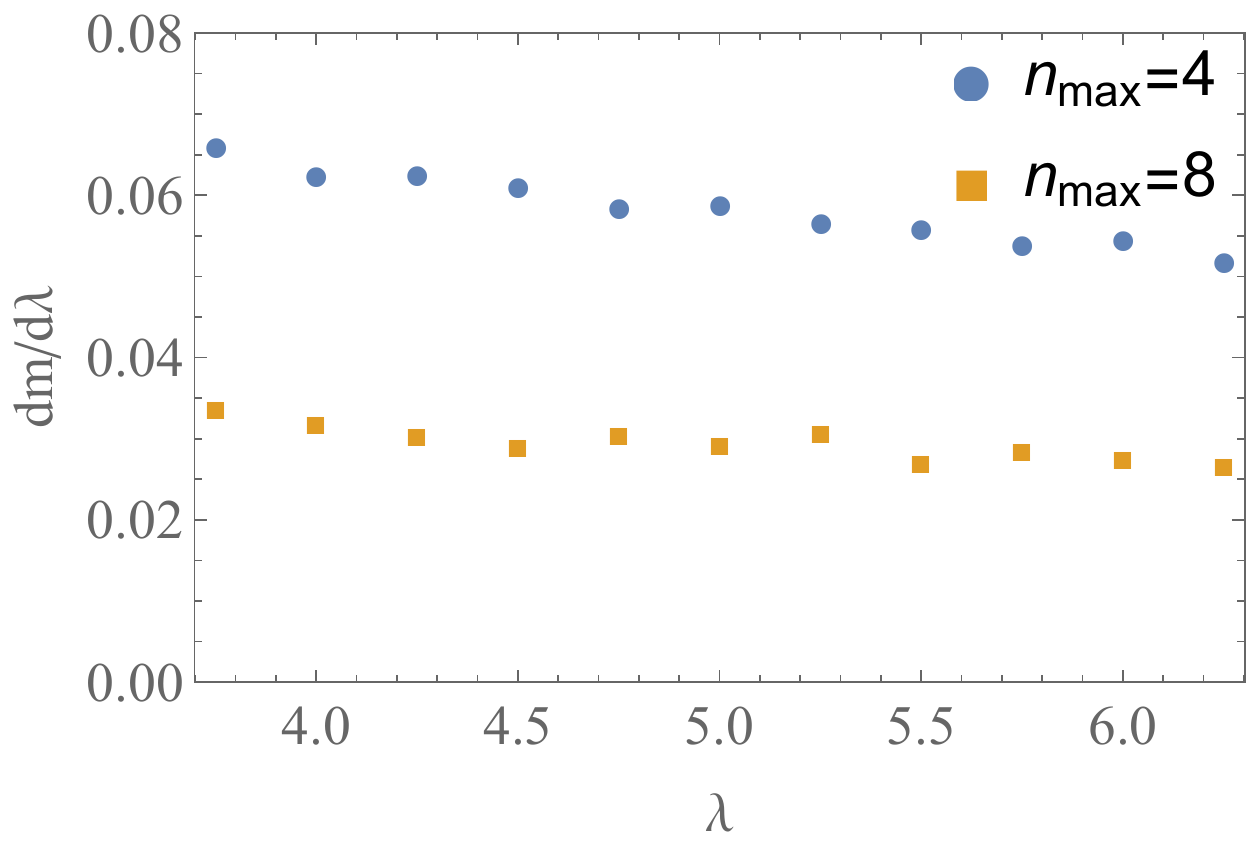}
	\includegraphics[scale=0.65]{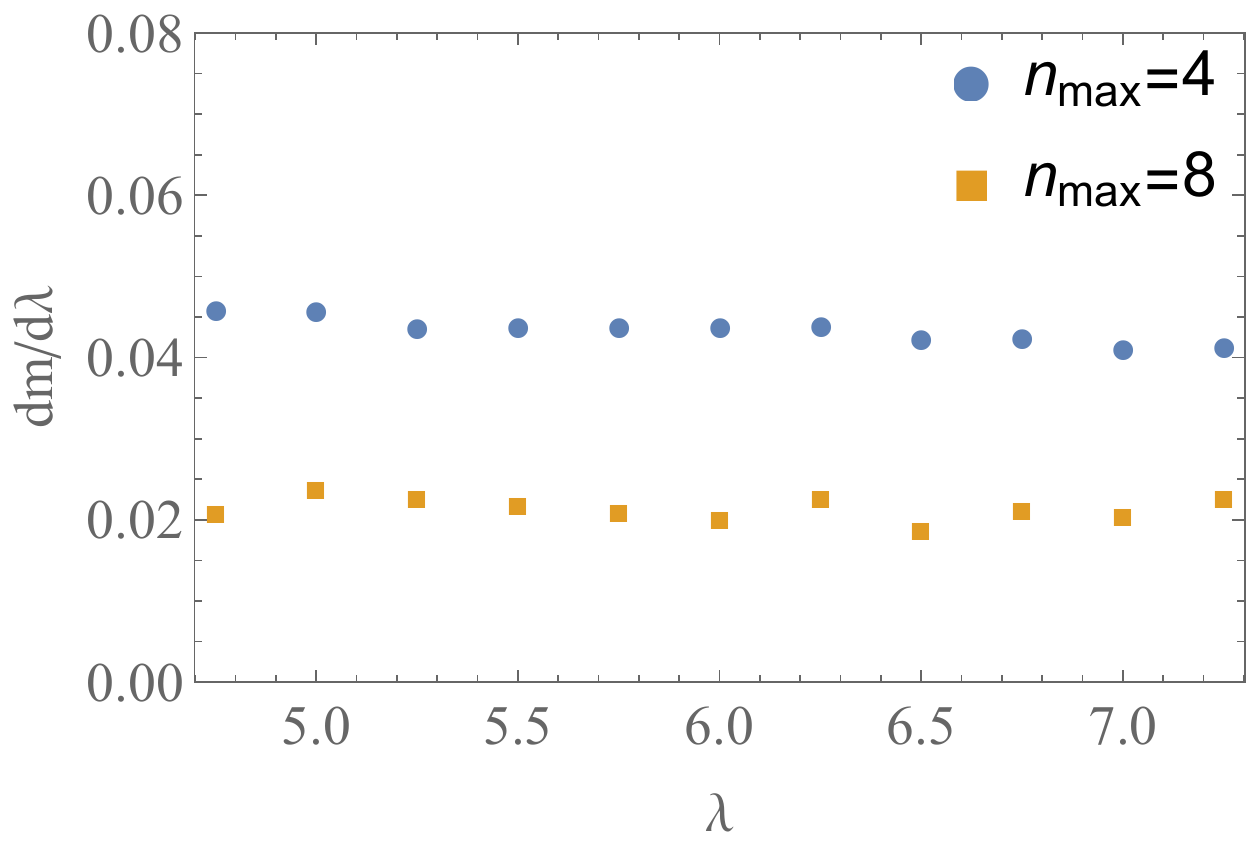}
	\end{center}
\caption{The dependence of the gap $m$ on the interaction coefficient $\lambda$, for $L=2$, Hilbert space cutoff $n_{\text{max}} = 4, 8$, and bare mass parameters $m_0^2=-1.5, -2.5$ (upper and lower panel, respectively) as $m\to 0$ showing that $\frac{dm}{d\lambda}$ approaches a constant at criticality. \label{critical_expo_above_lc_8}}
\end{figure}
As the system approaches the phase transition, it is expected to obey a power-law near the critical line. It undergoes a second-order phase transition such that, for a fixed value of the bare mass, $m_0^2$,
\be\label{eq:crit}
m \sim |\lambda-\lambda_c|^\nu~,
\ee
where $\lambda_c$ is the value of the coupling constant at the critical line.
Generally, critical properties of systems are independent of their microscopic structure depending only on the dimensionality of the system and the universality class the system belongs to. The critical exponent, $\nu$, which is a critical property of our scalar QFT, depends only on the dimension of space, which is 1 in our case. $\phi^4$ scalar field theory is expected to be in the same universality class as the Ising model (see ref.\ \cite{Jordan:2011ci} and references therein). 

The critical exponent for the Ising Model in one spatial dimension is $\nu=1$. The results of our analysis of the mass gap near the phase transition are consistent with the expected $\nu$ value for one spatial dimension. As we approach the critical line from above, keeping $m_0^2$ constant, approximations to $\lambda_c$ values are seen in Figure \ref{critical_expo_above_lc_8}. Moreover, the derivative $dm/d\lambda$ approaches a constant (albeit noisily, due to the low Hilbert space cutoff chosen, $n_{\text{max}} = 4,8$) as $\lambda \to \lambda_c^+$, providing evidence that near (and above) the critical line, the mass gap is a linear function of $\lambda$, supporting the expectation that the critical exponent (Eq.\ \eqref{eq:crit}) is $\nu = 1$.

\begin{figure*}[ht!]
\includegraphics[width=2\columnwidth]{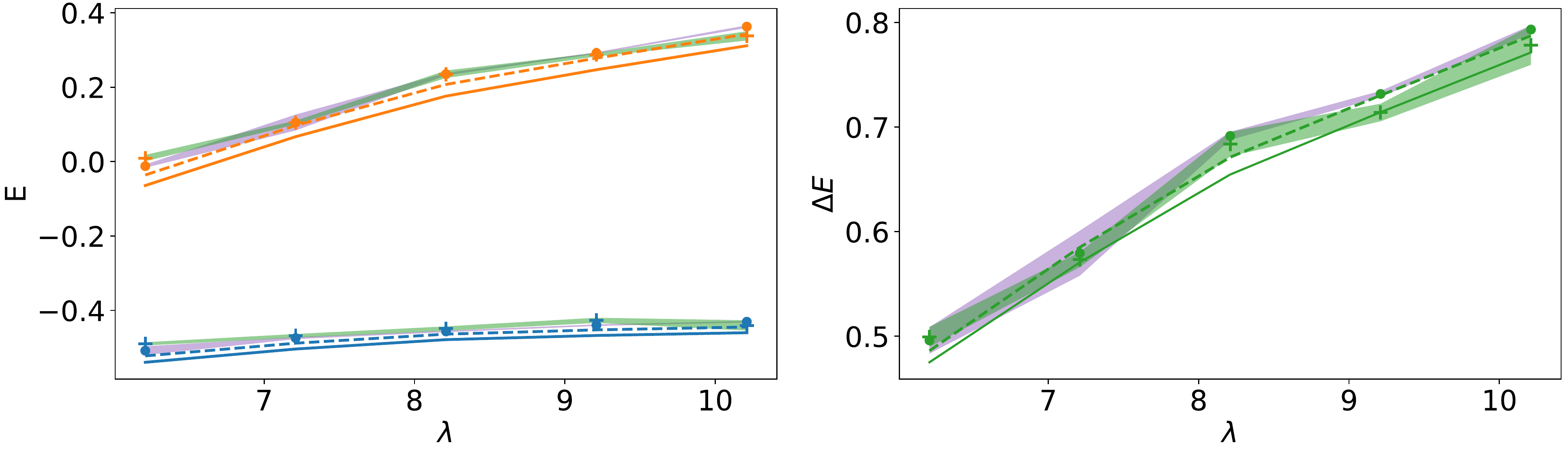}
\caption{Experimentally optimized ground and lowest lying excited energies as a function of the interaction parameter $\lambda$ are given in the left panel. Experimentally determined mass gaps $\Delta E=E_1-E_0$ are given in the right panel. The dashed (solid) lines correspond to the theoretically optimal product (entangled) ansatz energies, which are obtained by noiseless numerical simulation. Experimental results for the product (entangled) ansatz are given by the $\bullet$ ($+$) markers, where purple (green) error bars indicating one standard deviation are determined by repeated sampling of the energy functional at the optimal parameters. For all data points, we have taken $m_0^{2} =-1.5$ with $\delta_m$ being determined by Eq.~\eqref{eq:3}.}
\label{exp_data}
\end{figure*}

\section{Quantum Simulations}
\label{sec:sim}
With the theoretical description and numerical simulations completed, we can now compare the accuracy of quantum simulations and use this metric to judge the quantum computer's performance in QFT calculations. The calculations were performed on IBM's Tokyo chip, a device consisting of 20 fixed frequency transmon qubits \cite{Koch2007}. 
\subsection{Field Digitization}
\label{ssec:encoding}

We first impose periodic boundary conditions and consider the model system in Sec.~\ref{sec:set} consisting of $L=2$ spatial points. Next, the free part of Eq.~\eqref{eq:H} is diagonalized by Fourier transformation, with the symmetric and anti-symmetric momentum states $k \in \left\{0,\pi\right\}$ being the solutions to the free theory. Each momentum oscillator $\phi(k)$ still contains a real continuous degree of freedom, which would ideally be modeled by devices which can encode quantum information into continuous variables~\cite{Marshall:2015mna,Gao2018}. However, the public availability of such platforms remains limited, and this motivates us to encode the problem directly into a collection of two-level systems. In this case, we digitize each continuous degree of freedom in a truncated Fock space of the momentum modes as discussed in Sec.~\ref{subsec:Class_Sim}. By constraining the simulation to the truncated low-energy Fock space, i.e., up to a cutoff occupancy $n_{\text{max}}$, we only require $\mathcal{O}(\log_2(n_{\text{max}}))$ qubits per field mode. While a real space encoding and discretization is more appropriate in the asymptotic limit \cite{Klco2018}, the momentum space encoding remains amenable due to the simplicity of the interaction term for the special case of a two site system. 

Let us now outline the encoding scheme used to evaluate the Hamiltonian defined in Section~\ref{sec:set}. The creation and annihilation ladder operators for each mode are defined  in terms of their matrix elements $a^\dagger(k) = \sum_{n=0}^\infty \sqrt{n+1} \ket{n+1}\bra{n}_k$, and $a(k) = \sum_{n=1}^\infty \sqrt{n} \ket{n-1}\bra{n}_k$. The number operator is simply $n(k) = a^\dagger(k)a(k) = \sum_{n=0}^\infty n \ket{n}\bra{n}_k$. Each of these operators, and higher order functions of them, may be transformed into qubit spin-operators by a direct matrix element encoding. To encode the raising and lowering operators, note that they can be written in the general form $O = \sum_{i} \alpha_i \ket{i}\bra{j}$, where $j\in(i-1,i,i+1)$, such that the operator consists of the set of diagonal, super-diagonal, or sub-diagonal matrix elements. Regardless of the exact form, we consider an encoding of $\ket{i}\bra{j}$ in the computational basis $\ket{b_i}\bra{b_j}$, that is, given by the binary vector expansion $b_i = (b^{(i)}_0,b^{(i)}_1,...,b^{(i)}_{n_q-1})$ for the $n_q$ qubit register encoding each oscillator mode. Each of the composite matrix elements can be factorized into the tensor product of single qubit matrix elements as $\ket{i}\bra{j} = \bigotimes_l \ket{b^{(i)}_l} \bra{b^{(j)}_l}$. The four possibilities for the $l$-th component can be expressed in the Pauli basis as
\bea
\ket{0}\bra{0}_l & \rightarrow & \frac{\mathbb{I}+Z_l}{2}, \;\; \ket{0}\bra{1}_l  \rightarrow  \frac{X_l+iY_l}{2},  \nonumber \\
\ket{1}\bra{0}_l  & \rightarrow & \frac{X_l-iY_l}{2}, \;\;  \ket{1}\bra{1}_l \rightarrow \frac{\mathbb{I}-Z_l}{2}.
\eea
Examples of this matrix element encoding for the case of a cutoff $n_\text{max}=7$ can be seen in ref.~\cite{Klco2018}, Eqs.~(23) and (24).
In our quantum simulations we cutoff the local Fock spaces at  $n_{\text{max}}=3$ with the binary encoding $\ket{b_i} \Leftrightarrow \ket{i}$ mapping the truncated space as $\{\ket{00}, \ket{01}, \ket{10}, \ket{11}\} \Leftrightarrow \{\ket{0},\ket{1},\ket{2},\ket{3}\}$. 

The choice of cutoff (or field discretization) is not only a source for algorithmic error (see Fig.~\ref{fig:GS2} for this dependence) but it also dictates the circuit complexity, by controlling the number of variational parameters, ansatz depth, and other characteristics of the quantum program. Without error correction physical errors also accumulate rapidly and additional computations are required to improve the accuracy of the results. The trade-off between algorithmic error and physical hardware errors is an open field of study, but an optimal implementation resulting in low algorithmic and physical errors ostensibly exists for each algorithm~\cite{Endo2018, Hamilton2018}. Therefore, the maximum cutoff -- which determines the size of the quantum algorithm -- for which accurate results are attainable is a suitable metric across NISQ devices.

As discussed in Sec.~\ref{sec:mod}, the counter and interaction terms appearing in $H_I$ make the total Hamiltonian non-trivial. 
Defining $q(k) =\frac{1}{\sqrt{2}} \left[a(k) + a^\dagger(k)\right]$, we can write the discretized interaction Hamiltonian for $L=2$ as
\bea
H_I&=& \frac{\lambda}{48}\left[ \frac{q^4(0)}{\omega^2(0)} +\frac{6 q^2(0) q^2(\pi)}{\omega(0)\omega(\pi)}+\frac{q^4(\pi)}{\omega^2(\pi)} \right] \nonumber\\ 
& & + \frac{\delta_m}{2}\left[ \frac{q^2(0)}{\omega(0)}+ \frac{q^2(\pi)}{\omega(\pi)}\right]  \label{int_Ham}
\eea
Note that $q^2(k)=\frac{1}{2}\left[a^2(k) + (a^\dagger (k))^2 + 2n(k) + 1\right]$ so that, aside from the diagonal number operator, $H_I$ only connects occupation levels differing by $\pm2$. Since Eq.~\eqref{int_Ham} involves the sum over products of even powers of $q(k)$ there exists a global $\mathbb{Z}_2\times\mathbb{Z}_2$ symmetry decoupling the system into 4 distinct parity sectors. That is, the global parity sectors are constructed by the tensor product of local even/odd parity sectors. We take advantage of this symmetry and decouple the ground and first excited state subspaces. That is, the ground (excited) state belongs to the $\left\{+,+\right\}$($\left\{-,+\right\}$) parity block. Parity blocking the space allows us to represent each local mode with a single qubit, which now either encodes the even $\{\ket{0},\ket{2}\}$ or odd $\{\ket{1}, \ket{3}\}$ subspaces.

\subsection{Quantum Program}
We parametrize the hardware's configuration space with a pair of state preperation circuits (ansatze). The product instruction set is simply a pair of local rotations; $U_p = R_{Y_0} (\theta_0) R_{Y_1} (\theta_1)$ where $R_{Y_i} (\theta) \equiv e^{-i\frac{\theta}{2}Y_i}$ is the familiar rotation generated by the Pauli $Y$ operator on the $i$th qubit. $U_p$ is used to span the real amplitude tensor product subspace. 
With real coefficients sufficing to describe the eigenstates of the real symmetric parent Hamiltonian (Eq.~\eqref{eq:H}).
Of course, the system's eigenstates are not product states, but require entanglement. Adding a pair of controlled not (CNOT) gates we write the entangled ansatz $U_e = C_0R_{Y_1}(\theta_2) U_p$, where $C_iR_{\sigma^\mu_j}(\theta)$ denotes a controlled rotation of $\theta$ about the $\sigma^\mu_j$-axis contingent on the state of the $i$th qubit. For our minimal construction, the true eigenstates lie within the manifold spanned by $U_e$.
We implement $C_iR_{Y_j}(\theta)$ on hardware by expressing it as $R_{Y_j}(\frac{\theta}{2}) C_iX_j R_{Y_j}(\frac{-\theta}{2}) C_iX_j$\cite{QC_Bible}, with $i$ ($j$) denoting the control (target) CNOT qubits. 
Given that the energy difference between the optimal product and entangled states is small, about 1\%, the potential for the entangled ansatz to improve the renormalized mass depends strongly on the amount of noise, which is dominated by CNOT gates in our experiment. We must therefore employ error mitigation protocols within the classical component of our hybrid computation. 

\subsection{Error Mitigation}
To increase the precision of the noisy quantum simulator, error mitigating techniques must be applied to the raw experimental data. We begin two-stage error mitigation by first correcting for readout (RO) errors using a local (i.e. spatially uncorrelated) readout error model. We do so by first determining the rates at which individual qubits are flipped during RO with a set of pre-processing circuits. To estimate the individual bitflip rates each qubit (we map to qubits (0,1) and (0,11) for the entangled and product circuits respectively) is prepared in both computational basis states and immediately measured. The maximum likelihood estimates for the $i$th qubit flipping from $\ket{0}$ to $\ket{1}$ (and vice versa) is denoted as $p_i(0|1) (p_i(1|0))$. Denoting the symmetric and anti-symmetric combinations as $p_i^\pm = p(0|1) \pm p(1|0)$, RO-corrected expectation values are approximated as
\be
\label{eq:fix}
\expect{Z\cdots Z} = \sum_{x\in \text{counts}} p(x) \prod_{i \in \text{supp}(\expect{Z\cdots Z})}\left[ \frac{(-1)^{x_i} - p_i^-}{1-p_i^+}\right],
\ee
where supp($\expect{Z\cdots Z}$) refers to the operator support, i.e., the set of qubits $Z\cdots Z$ acts upon. While such a model may not capture correlation effects, this protocol scales linearly with the simulator system size \cite{kandala2017, Dumitrescu2018}. 

While RO mitigation is sufficient for $U_p$, we must further mitigate the errors induced by the noisy CNOTs in $U_e$. Recall that the variational principle ensures that energies estimated from variational methods upper bound the true ground state which is the eigenvalue corresponding lowest lying eigenvector, a pure state. On the other hand, states produced by  NISQ hardware are statistical mixtures of many pure states. To deal with this issue we employ an error mitigation scheme purifying our estimator quantum state at each step of classical optimization. First we independently measure the raw experimental state along each element of the Pauli group $\mathbb{P}_2 = \sigma^\mu_0\sigma^\nu_1$ for $\mu,\nu \in \{I,X,Y,Z\}$ writing $\rho = \sum_{p\in \mathbb{P}_2} c_p p$. We then purify\cite{Truflandier2016} the reference state by iterative updating the density matrix as $\rho_{n+1} = 3 \rho^2 - 2 \rho^3$ until the state's non-idempotency, defined as $\mathcal{N} = \text{Tr}\left[ \rho^2-\rho\right]$, lies below a threshold $\epsilon_\mathcal{N}<10^{-4}$. 

\subsection{Results and Discussion}
Panel (a) of Fig.~\ref{exp_data} shows the error corrected ground (and first excited) state energy coming from the $\left\{+,+\right\}$ ($\left\{-,+\right\}$) Hamiltonian parity blocks. In all simulations we have chosen a bare mass of $m_0^{2} =-1.5$ with $\delta_m$ determined by Eq.~\ref{eq:3}. Panel (b) illustrates the mass gap calculated by the difference $\Delta E = E_1-E_0$. The low-energy spectrum and gap are found to be in relatively close agreement with theory, i.e. the mass gap lies within a single standard deviation aside from the $\lambda=8.21$ points. The linear dependence of the spectral gap on the interaction coefficient is also qualitatively reproduced. 

These results demonstrate that a modern quantum computer can simulate the basic ingredients of a mass renormalization for a simple QFT. While the entangled ansatz only narrowly outperforms the product ansatz, the path to more complex and accurate QFT simulations is straightforward: higher cutoffs, with commensurately reduced CNOT noise, would allow more accurate simulations with larger lattice sizes. In general, more resources, in the form of lower-noise two qubit gates, are needed to go beyond $L=2$. The number of qubits required is $L\times \log_2 n_{\text{max}}$. The complexity of the algorithm increases with the cutoff $n_{\text{max}}$ and the number of points $L$, as well with the quantum circuit needed to generate an appropriate trial state. 

\section{Conclusion}
In this work we have performed a first exploration of the practical challenges associated with simulating an interacting scalar QFT on a NISQ device. To do so, we have developed a minimal construction and simulated it with a hybrid classical-quantum program. In order to obtain accurate results in the presence of noise, we strive to balance algorithmic and device errors in our simulation workflow. We find the entangled states slightly outperform, given a suitably robust error mitigation strategy, the product circuit in terms of mass gap accuracy. However, the theoretical separation between the product and entangled mass gaps was often larger than the observed statistical errors, but this difference was not discernible due to systematic device noise. 

While more complex simulations, with larger cutoff or a greater number of spatial sites, could be implemented on today's hardware, the results would likely not result in the same levels of accuracy. This loss of accuracy would be due to an increase in device noise, via a larger required circuit depth, which would mask any algorithmic error reduction. However, the path to improving QFT simulations on future hardware is clear: alleviating the bottlenecks with improved two qubit gates would allow one to increase both $L$ and $n_{max}$ while retaining or improving the final accuracy. Further, for larger system sizes alternative encodings, e.g., in a spatial basis, can reduce the overall algorithm complexity. 

Overall, hybrid algorithms are a viable means to simulate QFTs in the pre-fault-tolerant era and our results delineate the simulable and non-simulable problem sizes, mainly in terms of circuit complexity. In this sense, the algorithm presented here benchmarks NISQ computers' ability to  accurately simulate QFTs. The results given in terms of two different state preperation circuits point directly to which hardware improvements will enable increasingly complex (and accurate) algorithms. Future studies should involve different encodings, larger lattice sizes, alternative error mitigation strategies, and machine learning techniques in order to scale hybrid programs to meaningful scientific problems. 

\label{sec:con}

\acknowledgments
We acknowledge useful discussions with M. Savage and N. Klco and T. Morris. K.\ Y.\ A.\ was supported in part by an appointment to the Oak Ridge National Laboratory HERE Faculty Program, sponsored by the U.S.\ Department of Energy and administered by the Oak Ridge Institute for Science and Education.
E.\ F.\ D., A.\ J.\ M., and R.\ C.\ P.\, and K.~Y.~A.\ acknowledge DOE ASCR funding under the Testbed Pathfinder program, FWP number ERKJ332.
G.\ S.\ acknowledges support from the U.S.\ Office of Naval Research under award number N00014-15-1-2646. The authors acknowledge use of the IBM Q for this work. The views expressed are those of the authors and do not reflect the official policy or position of IBM or the IBM Q team.

%


\begin{thebibliography}{23}%
\makeatletter
\providecommand \@ifxundefined [1]{%
 \@ifx{#1\undefined}
}%
\providecommand \@ifnum [1]{%
 \ifnum #1\expandafter \@firstoftwo
 \else \expandafter \@secondoftwo
 \fi
}%
\providecommand \@ifx [1]{%
 \ifx #1\expandafter \@firstoftwo
 \else \expandafter \@secondoftwo
 \fi
}%
\providecommand \natexlab [1]{#1}%
\providecommand \enquote  [1]{``#1''}%
\providecommand \bibnamefont  [1]{#1}%
\providecommand \bibfnamefont [1]{#1}%
\providecommand \citenamefont [1]{#1}%
\providecommand \href@noop [0]{\@secondoftwo}%
\providecommand \href [0]{\begingroup \@sanitize@url \@href}%
\providecommand \@href[1]{\@@startlink{#1}\@@href}%
\providecommand \@@href[1]{\endgroup#1\@@endlink}%
\providecommand \@sanitize@url [0]{\catcode `\\12\catcode `\$12\catcode
  `\&12\catcode `\#12\catcode `\^12\catcode `\_12\catcode `\%12\relax}%
\providecommand \@@startlink[1]{}%
\providecommand \@@endlink[0]{}%
\providecommand \url  [0]{\begingroup\@sanitize@url \@url }%
\providecommand \@url [1]{\endgroup\@href {#1}{\urlprefix }}%
\providecommand \urlprefix  [0]{URL }%
\providecommand \Eprint [0]{\href }%
\providecommand \doibase [0]{http://dx.doi.org/}%
\providecommand \selectlanguage [0]{\@gobble}%
\providecommand \bibinfo  [0]{\@secondoftwo}%
\providecommand \bibfield  [0]{\@secondoftwo}%
\providecommand \translation [1]{[#1]}%
\providecommand \BibitemOpen [0]{}%
\providecommand \bibitemStop [0]{}%
\providecommand \bibitemNoStop [0]{.\EOS\space}%
\providecommand \EOS [0]{\spacefactor3000\relax}%
\providecommand \BibitemShut  [1]{\csname bibitem#1\endcsname}%
\let\auto@bib@innerbib\@empty
\bibitem [{\citenamefont {Feynman}(1982)}]{feynman_simulating_1982}%
  \BibitemOpen
  \bibfield  {author} {\bibinfo {author} {\bibfnamefont {R.~P.}\ \bibnamefont
  {Feynman}},\ }\href {\doibase 10.1007/BF02650179} {\bibfield  {journal}
  {\bibinfo  {journal} {International Journal of Theoretical Physics}\ }\textbf
  {\bibinfo {volume} {21}},\ \bibinfo {pages} {467} (\bibinfo {year}
  {1982})}\BibitemShut {NoStop}%
\bibitem [{\citenamefont {Lloyd}(1996)}]{Lloyd1996}%
  \BibitemOpen
  \bibfield  {author} {\bibinfo {author} {\bibfnamefont {S.}~\bibnamefont
  {Lloyd}},\ }\href {\doibase 10.1126/science.273.5278.1073} {\bibfield
  {journal} {\bibinfo  {journal} {Science}\ }\textbf {\bibinfo {volume}
  {273}},\ \bibinfo {pages} {1073?1078} (\bibinfo {year} {1996})}\BibitemShut
  {NoStop}%
\bibitem [{\citenamefont {Jordan}\ \emph {et~al.}(2012)\citenamefont {Jordan},
  \citenamefont {Lee},\ and\ \citenamefont {Preskill}}]{JordanLeePreskill2012}%
  \BibitemOpen
  \bibfield  {author} {\bibinfo {author} {\bibfnamefont {S.~P.}\ \bibnamefont
  {Jordan}}, \bibinfo {author} {\bibfnamefont {K.~S.~M.}\ \bibnamefont {Lee}},
  \ and\ \bibinfo {author} {\bibfnamefont {J.}~\bibnamefont {Preskill}},\
  }\href {\doibase 10.1126/science.1217069} {\bibfield  {journal} {\bibinfo
  {journal} {Science}\ }\textbf {\bibinfo {volume} {336}},\ \bibinfo {pages}
  {1130} (\bibinfo {year} {2012})}\BibitemShut {NoStop}%
\bibitem [{\citenamefont {Marshall}\ \emph {et~al.}(2015)\citenamefont
  {Marshall}, \citenamefont {Pooser}, \citenamefont {Siopsis},\ and\
  \citenamefont {Weedbrook}}]{Marshall:2015mna}%
  \BibitemOpen
  \bibfield  {author} {\bibinfo {author} {\bibfnamefont {K.}~\bibnamefont
  {Marshall}}, \bibinfo {author} {\bibfnamefont {R.}~\bibnamefont {Pooser}},
  \bibinfo {author} {\bibfnamefont {G.}~\bibnamefont {Siopsis}}, \ and\
  \bibinfo {author} {\bibfnamefont {C.}~\bibnamefont {Weedbrook}},\ }\href
  {\doibase 10.1103/PhysRevA.92.063825} {\bibfield  {journal} {\bibinfo
  {journal} {Phys. Rev. A}\ }\textbf {\bibinfo {volume} {92}},\ \bibinfo
  {pages} {063825} (\bibinfo {year} {2015})}\BibitemShut {NoStop}%
\bibitem [{\citenamefont {Preskill}(2018)}]{Preskill2018quantumcomputingin}%
  \BibitemOpen
  \bibfield  {author} {\bibinfo {author} {\bibfnamefont {J.}~\bibnamefont
  {Preskill}},\ }\href {\doibase 10.22331/q-2018-08-06-79} {\bibfield
  {journal} {\bibinfo  {journal} {{Quantum}}\ }\textbf {\bibinfo {volume}
  {2}},\ \bibinfo {pages} {79} (\bibinfo {year} {2018})}\BibitemShut {NoStop}%
\bibitem [{\citenamefont {Peruzzo}\ \emph {et~al.}(2014)\citenamefont
  {Peruzzo}, \citenamefont {McClean}, \citenamefont {Shadbolt}, \citenamefont
  {Yung}, \citenamefont {Zhou}, \citenamefont {Love}, \citenamefont
  {Aspuru-Guzik},\ and\ \citenamefont {O?Brien}}]{peruzzo_variational_2014}%
  \BibitemOpen
  \bibfield  {author} {\bibinfo {author} {\bibfnamefont {A.}~\bibnamefont
  {Peruzzo}}, \bibinfo {author} {\bibfnamefont {J.}~\bibnamefont {McClean}},
  \bibinfo {author} {\bibfnamefont {P.}~\bibnamefont {Shadbolt}}, \bibinfo
  {author} {\bibfnamefont {M.-H.}\ \bibnamefont {Yung}}, \bibinfo {author}
  {\bibfnamefont {X.-Q.}\ \bibnamefont {Zhou}}, \bibinfo {author}
  {\bibfnamefont {P.~J.}\ \bibnamefont {Love}}, \bibinfo {author}
  {\bibfnamefont {A.}~\bibnamefont {Aspuru-Guzik}}, \ and\ \bibinfo {author}
  {\bibfnamefont {J.~L.}\ \bibnamefont {O?Brien}},\ }\href {\doibase
  10.1038/ncomms5213} {\bibfield  {journal} {\bibinfo  {journal} {Nature
  Communications}\ }\textbf {\bibinfo {volume} {5}},\ \bibinfo {pages} {4213}
  (\bibinfo {year} {2014})}\BibitemShut {NoStop}%
\bibitem [{\citenamefont {Arrazola}\ \emph {et~al.}(2016)\citenamefont
  {Arrazola}, \citenamefont {Pedernales}, \citenamefont {Lamata},\ and\
  \citenamefont {Solano}}]{arrazola_digital-analog_2016}%
  \BibitemOpen
  \bibfield  {author} {\bibinfo {author} {\bibfnamefont {I.}~\bibnamefont
  {Arrazola}}, \bibinfo {author} {\bibfnamefont {J.~S.}\ \bibnamefont
  {Pedernales}}, \bibinfo {author} {\bibfnamefont {L.}~\bibnamefont {Lamata}},
  \ and\ \bibinfo {author} {\bibfnamefont {E.}~\bibnamefont {Solano}},\ }\href
  {\doibase 10.1038/srep30534} {\bibfield  {journal} {\bibinfo  {journal}
  {Scientific Reports}\ }\textbf {\bibinfo {volume} {6}} (\bibinfo {year}
  {2016}),\ 10.1038/srep30534}\BibitemShut {NoStop}%
\bibitem [{\citenamefont {Lloyd}\ and\ \citenamefont
  {Braunstein}(1999)}]{lloyd_quantum_1999}%
  \BibitemOpen
  \bibfield  {author} {\bibinfo {author} {\bibfnamefont {S.}~\bibnamefont
  {Lloyd}}\ and\ \bibinfo {author} {\bibfnamefont {S.~L.}\ \bibnamefont
  {Braunstein}},\ }\href {\doibase 10.1103/PhysRevLett.82.1784} {\bibfield
  {journal} {\bibinfo  {journal} {Physical Review Letters}\ }\textbf {\bibinfo
  {volume} {82}},\ \bibinfo {pages} {1784} (\bibinfo {year}
  {1999})}\BibitemShut {NoStop}%
\bibitem [{\citenamefont {Martinez}\ \emph {et~al.}(2016)\citenamefont
  {Martinez}, \citenamefont {Muschik}, \citenamefont {Schindler}, \citenamefont
  {Nigg}, \citenamefont {Erhard}, \citenamefont {Heyl}, \citenamefont {Hauke},
  \citenamefont {Dalmonte}, \citenamefont {Monz}, \citenamefont {Zoller},\ and\
  \citenamefont {Blatt}}]{Martinez2016}%
  \BibitemOpen
  \bibfield  {author} {\bibinfo {author} {\bibfnamefont {E.~A.}\ \bibnamefont
  {Martinez}}, \bibinfo {author} {\bibfnamefont {C.~A.}\ \bibnamefont
  {Muschik}}, \bibinfo {author} {\bibfnamefont {P.}~\bibnamefont {Schindler}},
  \bibinfo {author} {\bibfnamefont {D.}~\bibnamefont {Nigg}}, \bibinfo {author}
  {\bibfnamefont {A.}~\bibnamefont {Erhard}}, \bibinfo {author} {\bibfnamefont
  {M.}~\bibnamefont {Heyl}}, \bibinfo {author} {\bibfnamefont {P.}~\bibnamefont
  {Hauke}}, \bibinfo {author} {\bibfnamefont {M.}~\bibnamefont {Dalmonte}},
  \bibinfo {author} {\bibfnamefont {T.}~\bibnamefont {Monz}}, \bibinfo {author}
  {\bibfnamefont {P.}~\bibnamefont {Zoller}}, \ and\ \bibinfo {author}
  {\bibfnamefont {R.}~\bibnamefont {Blatt}},\ }\href
  {http://dx.doi.org/10.1038/nature18318} {\bibfield  {journal} {\bibinfo
  {journal} {Nature}\ }\textbf {\bibinfo {volume} {534}},\ \bibinfo {pages}
  {516} (\bibinfo {year} {2016})}\BibitemShut {NoStop}%
\bibitem [{\citenamefont {Clements}\ \emph {et~al.}(2017)\citenamefont
  {Clements}, \citenamefont {Renema}, \citenamefont {Eckstein}, \citenamefont
  {Valido}, \citenamefont {Lita}, \citenamefont {Gerrits}, \citenamefont {Nam},
  \citenamefont {Kolthammer}, \citenamefont {Huh},\ and\ \citenamefont
  {Walmsley}}]{clements_approximating_2017}%
  \BibitemOpen
  \bibfield  {author} {\bibinfo {author} {\bibfnamefont {W.~R.}\ \bibnamefont
  {Clements}}, \bibinfo {author} {\bibfnamefont {J.~J.}\ \bibnamefont
  {Renema}}, \bibinfo {author} {\bibfnamefont {A.}~\bibnamefont {Eckstein}},
  \bibinfo {author} {\bibfnamefont {A.~A.}\ \bibnamefont {Valido}}, \bibinfo
  {author} {\bibfnamefont {A.}~\bibnamefont {Lita}}, \bibinfo {author}
  {\bibfnamefont {T.}~\bibnamefont {Gerrits}}, \bibinfo {author} {\bibfnamefont
  {S.~W.}\ \bibnamefont {Nam}}, \bibinfo {author} {\bibfnamefont {W.~S.}\
  \bibnamefont {Kolthammer}}, \bibinfo {author} {\bibfnamefont
  {J.}~\bibnamefont {Huh}}, \ and\ \bibinfo {author} {\bibfnamefont {I.~A.}\
  \bibnamefont {Walmsley}},\ }\href {http://arxiv.org/abs/1710.08655}
  {\bibfield  {journal} {\bibinfo  {journal} {arXiv:1710.08655 [quant-ph]}\ }
  (\bibinfo {year} {2017})},\ \bibinfo {note} {arXiv: 1710.08655}\BibitemShut
  {NoStop}%
\bibitem [{\citenamefont {Klco}\ \emph {et~al.}(2018)\citenamefont {Klco},
  \citenamefont {Dumitrescu}, \citenamefont {McCaskey}, \citenamefont {Morris},
  \citenamefont {Pooser}, \citenamefont {Sanz}, \citenamefont {Solano},
  \citenamefont {Lougovski},\ and\ \citenamefont
  {Savage}}]{klco_quantum-classical_2018}%
  \BibitemOpen
  \bibfield  {author} {\bibinfo {author} {\bibfnamefont {N.}~\bibnamefont
  {Klco}}, \bibinfo {author} {\bibfnamefont {E.~F.}\ \bibnamefont
  {Dumitrescu}}, \bibinfo {author} {\bibfnamefont {A.~J.}\ \bibnamefont
  {McCaskey}}, \bibinfo {author} {\bibfnamefont {T.~D.}\ \bibnamefont
  {Morris}}, \bibinfo {author} {\bibfnamefont {R.~C.}\ \bibnamefont {Pooser}},
  \bibinfo {author} {\bibfnamefont {M.}~\bibnamefont {Sanz}}, \bibinfo {author}
  {\bibfnamefont {E.}~\bibnamefont {Solano}}, \bibinfo {author} {\bibfnamefont
  {P.}~\bibnamefont {Lougovski}}, \ and\ \bibinfo {author} {\bibfnamefont
  {M.~J.}\ \bibnamefont {Savage}},\ }\href {\doibase
  10.1103/PhysRevA.98.032331} {\bibfield  {journal} {\bibinfo  {journal}
  {Physical Review A}\ }\textbf {\bibinfo {volume} {98}},\ \bibinfo {pages}
  {032331} (\bibinfo {year} {2018})},\ \Eprint
  {http://arxiv.org/abs/1803.03326} {arXiv:1803.03326} \BibitemShut {NoStop}%
\bibitem [{\citenamefont {Kokail}\ \emph {et~al.}(2018)\citenamefont {Kokail},
  \citenamefont {Maier}, \citenamefont {van Bijnen}, \citenamefont {Brydges},
  \citenamefont {Joshi}, \citenamefont {Jurcevic}, \citenamefont {Muschik},
  \citenamefont {Silvi}, \citenamefont {Blatt}, \citenamefont {Roos},\ and\
  \citenamefont {Zoller}}]{Kokail2018}%
  \BibitemOpen
  \bibfield  {author} {\bibinfo {author} {\bibfnamefont {C.}~\bibnamefont
  {Kokail}}, \bibinfo {author} {\bibfnamefont {C.}~\bibnamefont {Maier}},
  \bibinfo {author} {\bibfnamefont {R.}~\bibnamefont {van Bijnen}}, \bibinfo
  {author} {\bibfnamefont {T.}~\bibnamefont {Brydges}}, \bibinfo {author}
  {\bibfnamefont {M.~K.}\ \bibnamefont {Joshi}}, \bibinfo {author}
  {\bibfnamefont {P.}~\bibnamefont {Jurcevic}}, \bibinfo {author}
  {\bibfnamefont {C.~A.}\ \bibnamefont {Muschik}}, \bibinfo {author}
  {\bibfnamefont {P.}~\bibnamefont {Silvi}}, \bibinfo {author} {\bibfnamefont
  {R.}~\bibnamefont {Blatt}}, \bibinfo {author} {\bibfnamefont {C.~F.}\
  \bibnamefont {Roos}}, \ and\ \bibinfo {author} {\bibfnamefont
  {P.}~\bibnamefont {Zoller}},\ }\href {http://arxiv.org/abs/1810.03421} {\
  (\bibinfo {year} {2018})},\ \Eprint {http://arxiv.org/abs/1810.03421}
  {arXiv:1810.03421} \BibitemShut {NoStop}%
\bibitem [{noa(2018)}]{noauthor_quantum_2018}%
  \BibitemOpen
  \href {https://www.research.ibm.com/ibm-q/technology/devices/}
  {{\enquote {\bibinfo {title} {Quantum devices \&
  simulators},}\ }}\bibinfo {howpublished}
  {\url{https://www.research.ibm.com/ibm-q/technology/devices/}} (\bibinfo
  {year} {2018}),\ \bibinfo {note} {accessed: 2018-11-22}\BibitemShut {NoStop}%
\bibitem [{\citenamefont {Jordan}\ \emph {et~al.}(2011)\citenamefont {Jordan},
  \citenamefont {Lee},\ and\ \citenamefont {Preskill}}]{Jordan:2011ci}%
  \BibitemOpen
  \bibfield  {author} {\bibinfo {author} {\bibfnamefont {S.~P.}\ \bibnamefont
  {Jordan}}, \bibinfo {author} {\bibfnamefont {K.~S.~M.}\ \bibnamefont {Lee}},
  \ and\ \bibinfo {author} {\bibfnamefont {J.}~\bibnamefont {Preskill}},\
  }\href@noop {} {\  (\bibinfo {year} {2011})},\ \bibinfo {note} {[Quant. Inf.
  Comput.14,1014(2014)]},\ \Eprint {http://arxiv.org/abs/1112.4833}
  {arXiv:1112.4833 [hep-th]} \BibitemShut {NoStop}%
\bibitem [{\citenamefont {Koch}\ \emph {et~al.}(2007)\citenamefont {Koch},
  \citenamefont {Yu}, \citenamefont {Gambetta}, \citenamefont {Houck},
  \citenamefont {Schuster}, \citenamefont {Majer}, \citenamefont {Blais},
  \citenamefont {Devoret}, \citenamefont {Girvin},\ and\ \citenamefont
  {Schoelkopf}}]{Koch2007}%
  \BibitemOpen
  \bibfield  {author} {\bibinfo {author} {\bibfnamefont {J.}~\bibnamefont
  {Koch}}, \bibinfo {author} {\bibfnamefont {T.~M.}\ \bibnamefont {Yu}},
  \bibinfo {author} {\bibfnamefont {J.}~\bibnamefont {Gambetta}}, \bibinfo
  {author} {\bibfnamefont {A.~A.}\ \bibnamefont {Houck}}, \bibinfo {author}
  {\bibfnamefont {D.~I.}\ \bibnamefont {Schuster}}, \bibinfo {author}
  {\bibfnamefont {J.}~\bibnamefont {Majer}}, \bibinfo {author} {\bibfnamefont
  {A.}~\bibnamefont {Blais}}, \bibinfo {author} {\bibfnamefont {M.~H.}\
  \bibnamefont {Devoret}}, \bibinfo {author} {\bibfnamefont {S.~M.}\
  \bibnamefont {Girvin}}, \ and\ \bibinfo {author} {\bibfnamefont {R.~J.}\
  \bibnamefont {Schoelkopf}},\ }\href {\doibase 10.1103/PhysRevA.76.042319}
  {\bibfield  {journal} {\bibinfo  {journal} {Phys. Rev. A}\ }\textbf {\bibinfo
  {volume} {76}},\ \bibinfo {pages} {042319} (\bibinfo {year}
  {2007})}\BibitemShut {NoStop}%
\bibitem [{\citenamefont {Gao}\ \emph {et~al.}(2018)\citenamefont {Gao},
  \citenamefont {Lester}, \citenamefont {Chou}, \citenamefont {Frunzio},
  \citenamefont {Devoret}, \citenamefont {Jiang}, \citenamefont {Girvin},\ and\
  \citenamefont {Schoelkopf}}]{Gao2018}%
  \BibitemOpen
  \bibfield  {author} {\bibinfo {author} {\bibfnamefont {Y.~Y.}\ \bibnamefont
  {Gao}}, \bibinfo {author} {\bibfnamefont {B.~J.}\ \bibnamefont {Lester}},
  \bibinfo {author} {\bibfnamefont {K.}~\bibnamefont {Chou}}, \bibinfo {author}
  {\bibfnamefont {L.}~\bibnamefont {Frunzio}}, \bibinfo {author} {\bibfnamefont
  {M.~H.}\ \bibnamefont {Devoret}}, \bibinfo {author} {\bibfnamefont
  {L.}~\bibnamefont {Jiang}}, \bibinfo {author} {\bibfnamefont {S.~M.}\
  \bibnamefont {Girvin}}, \ and\ \bibinfo {author} {\bibfnamefont {R.~J.}\
  \bibnamefont {Schoelkopf}},\ }\href {http://arxiv.org/abs/1806.07401} {\
  (\bibinfo {year} {2018})},\ \Eprint {http://arxiv.org/abs/1806.07401}
  {arXiv:1806.07401} \BibitemShut {NoStop}%
\bibitem [{\citenamefont {Klco}\ and\ \citenamefont {Savage}(2018)}]{Klco2018}%
  \BibitemOpen
  \bibfield  {author} {\bibinfo {author} {\bibfnamefont {N.}~\bibnamefont
  {Klco}}\ and\ \bibinfo {author} {\bibfnamefont {M.}~\bibnamefont {Savage}},\
  }\href {http://arxiv.org/abs/1808.10378} {\bibfield  {journal} {\bibinfo
  {journal} {arxiv:1808.10378 [quant-ph]}\ } (\bibinfo {year} {2018})},\
  \Eprint {http://arxiv.org/abs/1808.10378} {arXiv:1808.10378} \BibitemShut
  {NoStop}%
\bibitem [{\citenamefont {Endo}\ \emph {et~al.}(2018)\citenamefont {Endo},
  \citenamefont {Zhao}, \citenamefont {Li}, \citenamefont {Benjamin},\ and\
  \citenamefont {Yuan}}]{Endo2018}%
  \BibitemOpen
  \bibfield  {author} {\bibinfo {author} {\bibfnamefont {S.}~\bibnamefont
  {Endo}}, \bibinfo {author} {\bibfnamefont {Q.}~\bibnamefont {Zhao}}, \bibinfo
  {author} {\bibfnamefont {Y.}~\bibnamefont {Li}}, \bibinfo {author}
  {\bibfnamefont {S.}~\bibnamefont {Benjamin}}, \ and\ \bibinfo {author}
  {\bibfnamefont {X.}~\bibnamefont {Yuan}},\ }\href
  {http://arxiv.org/abs/1808.03623} {\bibfield  {journal} {\bibinfo  {journal}
  {arxiv:1808.03623, [quant-ph]}\ } (\bibinfo {year} {2018})},\ \Eprint
  {http://arxiv.org/abs/1808.03623} {arXiv:1808.03623} \BibitemShut {NoStop}%
\bibitem [{\citenamefont {Hamilton}\ \emph {et~al.}(2018)\citenamefont
  {Hamilton}, \citenamefont {Dumitrescu},\ and\ \citenamefont
  {Pooser}}]{Hamilton2018}%
  \BibitemOpen
  \bibfield  {author} {\bibinfo {author} {\bibfnamefont {K.~E.}\ \bibnamefont
  {Hamilton}}, \bibinfo {author} {\bibfnamefont {E.~F.}\ \bibnamefont
  {Dumitrescu}}, \ and\ \bibinfo {author} {\bibfnamefont {R.~C.}\ \bibnamefont
  {Pooser}},\ }\href {http://arxiv.org/abs/1811.09905} {\bibfield  {journal}
  {\bibinfo  {journal} {arXiv:1811.09905 [quant-ph]}\ } (\bibinfo {year}
  {2018})},\ \Eprint {http://arxiv.org/abs/1811.09905} {arXiv:1811.09905}
  \BibitemShut {NoStop}%
\bibitem [{\citenamefont {Nielsen}\ and\ \citenamefont
  {Chuang}(2011)}]{QC_Bible}%
  \BibitemOpen
  \bibfield  {author} {\bibinfo {author} {\bibfnamefont {M.~A.}\ \bibnamefont
  {Nielsen}}\ and\ \bibinfo {author} {\bibfnamefont {I.~L.}\ \bibnamefont
  {Chuang}},\ }\href@noop {} {\emph {\bibinfo {title} {Quantum Computation and
  Quantum Information: 10th Anniversary Edition}}},\ \bibinfo {edition} {10th}\
  ed.\ (\bibinfo  {publisher} {Cambridge University Press},\ \bibinfo {address}
  {New York, NY, USA},\ \bibinfo {year} {2011})\BibitemShut {NoStop}%
\bibitem [{\citenamefont {{Kandala}}\ \emph {et~al.}(2017)\citenamefont
  {{Kandala}}, \citenamefont {{Mezzacapo}}, \citenamefont {{Temme}},
  \citenamefont {{Takita}}, \citenamefont {{Brink}}, \citenamefont {{Chow}},\
  and\ \citenamefont {{Gambetta}}}]{kandala2017}%
  \BibitemOpen
  \bibfield  {author} {\bibinfo {author} {\bibfnamefont {A.}~\bibnamefont
  {{Kandala}}}, \bibinfo {author} {\bibfnamefont {A.}~\bibnamefont
  {{Mezzacapo}}}, \bibinfo {author} {\bibfnamefont {K.}~\bibnamefont
  {{Temme}}}, \bibinfo {author} {\bibfnamefont {M.}~\bibnamefont {{Takita}}},
  \bibinfo {author} {\bibfnamefont {M.}~\bibnamefont {{Brink}}}, \bibinfo
  {author} {\bibfnamefont {J.~M.}\ \bibnamefont {{Chow}}}, \ and\ \bibinfo
  {author} {\bibfnamefont {J.~M.}\ \bibnamefont {{Gambetta}}},\ }\href
  {\doibase 10.1038/nature23879} {\bibfield  {journal} {\bibinfo  {journal}
  {\nat}\ }\textbf {\bibinfo {volume} {549}},\ \bibinfo {pages} {242} (\bibinfo
  {year} {2017})},\ \Eprint {http://arxiv.org/abs/1704.05018} {arXiv:1704.05018
  [quant-ph]} \BibitemShut {NoStop}%
\bibitem [{\citenamefont {Dumitrescu}\ \emph {et~al.}(2018)\citenamefont
  {Dumitrescu}, \citenamefont {McCaskey}, \citenamefont {Hagen}, \citenamefont
  {Jansen}, \citenamefont {Morris}, \citenamefont {Papenbrock}, \citenamefont
  {Pooser}, \citenamefont {Dean},\ and\ \citenamefont
  {Lougovski}}]{Dumitrescu2018}%
  \BibitemOpen
  \bibfield  {author} {\bibinfo {author} {\bibfnamefont {E.~F.}\ \bibnamefont
  {Dumitrescu}}, \bibinfo {author} {\bibfnamefont {A.~J.}\ \bibnamefont
  {McCaskey}}, \bibinfo {author} {\bibfnamefont {G.}~\bibnamefont {Hagen}},
  \bibinfo {author} {\bibfnamefont {G.~R.}\ \bibnamefont {Jansen}}, \bibinfo
  {author} {\bibfnamefont {T.~D.}\ \bibnamefont {Morris}}, \bibinfo {author}
  {\bibfnamefont {T.}~\bibnamefont {Papenbrock}}, \bibinfo {author}
  {\bibfnamefont {R.~C.}\ \bibnamefont {Pooser}}, \bibinfo {author}
  {\bibfnamefont {D.~J.}\ \bibnamefont {Dean}}, \ and\ \bibinfo {author}
  {\bibfnamefont {P.}~\bibnamefont {Lougovski}},\ }\href {\doibase
  10.1103/PhysRevLett.120.210501} {\bibfield  {journal} {\bibinfo  {journal}
  {Physical Review Letters}\ }\textbf {\bibinfo {volume} {120}},\ \bibinfo
  {pages} {210501} (\bibinfo {year} {2018})},\ \Eprint
  {http://arxiv.org/abs/1801.03897} {arXiv:1801.03897} \BibitemShut {NoStop}%
\bibitem [{\citenamefont {Truflandier}\ \emph {et~al.}(2016)\citenamefont
  {Truflandier}, \citenamefont {Dianzinga},\ and\ \citenamefont
  {Bowler}}]{Truflandier2016}%
  \BibitemOpen
  \bibfield  {author} {\bibinfo {author} {\bibfnamefont {L.~A.}\ \bibnamefont
  {Truflandier}}, \bibinfo {author} {\bibfnamefont {R.~M.}\ \bibnamefont
  {Dianzinga}}, \ and\ \bibinfo {author} {\bibfnamefont {D.~R.}\ \bibnamefont
  {Bowler}},\ }\href {\doibase 10.1063/1.4943213} {\bibfield  {journal}
  {\bibinfo  {journal} {Journal of Chemical Physics}\ }\textbf {\bibinfo
  {volume} {144}} (\bibinfo {year} {2016}),\ 10.1063/1.4943213},\ \Eprint
  {http://arxiv.org/abs/1512.07236} {arXiv:1512.07236} \BibitemShut {NoStop}%
\end{thebibliography}
\end{document}